\documentclass[showpacs,nofootinbib,10pt]{revtex4}
\usepackage{mathtools}
\usepackage[dvipsnames]{xcolor}
\usepackage{color}
\usepackage{graphicx}
\usepackage{graphicx,float}\usepackage[all]{xy}
\usepackage{amsmath,upgreek,tikz,bm}
 
\usepackage{amssymb}

\usepackage{textgreek}

\newcommand{\beq}{\begin{eqnarray}}
\newcommand{\eeq}{\end{eqnarray}}
\newcommand{\be}{\begin{equation}}
\newcommand{\ee}{\end{equation}}

\newcommand{\bea}{\begin{eqnarray}}
\newcommand{\eea}{\end{eqnarray}}
\newcommand{\bes}{\begin{subequations}}
\newcommand{\ees}{\end{subequations}}

\newcommand{\ba}{\begin{eqnarray}}
\newcommand{\ea}{\end{eqnarray}}
\newcommand\orcidrogerio{{\href{https://orcid.org/0000-0001-7848-5472}{\orcidicon}}}
\newcommand\orcidjulio{{\href{https://orcid.org/0000-0001-7405-8852}{\orcidicon}}}
\newcommand{\orcidicon}{%
	\begin{tikzpicture}
	\draw[lime, fill=lime] (0,0)
		circle [radius=0.16]
		node[white] {{\fontfamily{qag}\selectfont \tiny ID}};
	\draw[white, fill=white] (-0.0625,0.095)
		circle [radius=0.007];
	\end{tikzpicture}	\hspace{-2mm}
}

\usepackage[colorlinks,hyperindex,unicode]{hyperref} 
\definecolor{green1}{RGB}{0,128,0} 
\hypersetup{hidelinks,backref=true,pagebackref=true,hyperindex=true,colorlinks=true,breaklinks=true,urlcolor= blue}
\hypersetup{%
  colorlinks = true,
  linkcolor  = blue,
  citecolor = cyan,
}
\usepackage{bookmark,textgreek}
\usepackage{hyperref,color,xcolor}
\hypersetup{hidelinks,hyperindex=true,colorlinks=true,breaklinks=true,urlcolor= blue}
\hypersetup{%
  colorlinks = true,
  linkcolor  = blue
}
\begin{document}
\title{Near horizon thermodynamics of hairy black holes from gravitational decoupling}

\author{R. T. Cavalcanti\orcidrogerio\!\!}
\email{rogerio.cavalcanti@ufabc.edu.br} 
\affiliation{Centro de Matemática, Computação e Cognição, Universidade Federal do ABC, 09210-580, Santo Andr\'e, Brazil.\\ Departamento de Física, Universidade Estadual Paulista, Unesp, Guaratinguet\'{a}, 12516-410, Brazil}
\author{K. dos S. Alves}
\email{kelvin.santos@unesp.br} \affiliation{DFI, Universidade Estadual Paulista, Unesp, Guaratinguet\'{a}, 12516-410, Brazil}
\author{J. M. Hoff da Silva\orcidjulio\!\!}
\email{julio.hoff@unesp.br}
\affiliation{Departamento de Física, Universidade Estadual Paulista, Unesp, Guaratinguet\'{a}, 12516-410, Brazil}



\begin{abstract} {
The horizon structure and thermodynamics of hairy spherically symmetric black holes generated by the gravitational decoupling method are carefully investigated. The temperature and heat capacity of the black hole is determined, as well as how the hairy parameters affect the thermodynamics. It allows the analysis of the thermal stability and the possible existence of a remanent black hole. We also calculate the Hawking radiation corrected by the generalized uncertainty principle. For such we consider the emission of fermions and apply the tunneling method to the generalized Dirac equation. It shows that, despite the horizon location being the same of the Schwarzschild one for a suitable choice of parameters, the physical phenomena happening near the horizon of both black holes are qualitatively different. }
\end{abstract}



\maketitle

\section{Introduction}

Black hole physics plays a central role in contemporary research \cite{Cardoso:2019rvt}, at scales ranging from cosmology and astrophysics \cite{romero2013introduction} to elementary particle physics \cite{calmet2015quantum}. The emergence of observational proposals for investigating such objects was made possible after the classical results of Penrose, Hawking, Geroch, Israel, Carter and others, which put the geometry and physics of black hole solutions over solid basis.Among the central results, could be pinpointed: the classic definition of event horizon, the singularity theorems by Penrose, Hawking, and Geroch, the no-hair theorem, the cosmic censorship hypothesis, and the formation of trapped surfaces under generic conditions during gravitational collapse. Notably, strong observational evidence in astrophysics has been accumulating since the 1970s \cite{romero2013introduction}, including the ones that culminated in the nobel prize for Reinhard Genzel \cite{Gillessen:2008qv} and Andrea M. Ghez \cite{Ghez:2008ms}, the LIGO detections of gravitational waves \cite{Abbott:2016blz} and the image captured by the \textit{Event Horizon Telescope} collaboration \cite{EventHorizonTelescope:2019dse}. Black holes are closely linked to some of the most powerful processes known to science, such as the gravitational collapse of stars, active galactic nuclei, and the aforementioned gravitational waves emitted by binary black hole systems. Formally, black holes are characterized by the existence of horizons bounding two causally disconnected regions \cite{frolov2011introduction,Faraoni:2015ula}. Its most important feature is then not the existence of singularity, which has no support in established physics, but of an event horizon covering its interior. While the characterization of stationary and asymptotically flat hole horizons in general relativity is well known \cite{Wald:1984rg}, the physical understanding of the horizon nature of solutions beyond general relativity or non-stationary ones have generated extensive research in recent years \cite{Faraoni:2015ula,Ashtekar:2005ez,Gourgoulhon:2008pu,Chrusciel:2020fql}.  In the development of black hole mechanics and thermodynamics, which culminated in Bekenstein's second law of generalized thermodynamics  and Hawking radiation, the horizon also occupies a central role. However, the definition of event horizon inspired by stationary black holes turns out to be of little use in investigations about objects with less symmetric dynamics. This major obstacle manifests itself in the fact that to precisely define the event horizon would require knowledge of the entire spacetime history, which is obviously not physically achievable \cite{Ashtekar:2005ez}. In addition, there are the alternative gravitation theories to general relativity. This may include the existence of non-minimally curvature-coupled scalar fields or terms with higher order derivatives in the action, which has direct consequences for the uniqueness theorems of black hole type solutions in general relativity. In fact, the famous \textit{no hair theorem} is not preserved outside the domain of general relativity. Such solutions could lead to detectable effects in the vicinity of the astrophysical black hole horizon \cite{Babichev:2013cya, Cavalcanti:2016mbe}.

Finding physically relevant solutions of the Einstein field equations is not an easy task. However, deriving new solutions from other previously known is widespread technique in general relativity. Recently the so called gravitational decoupling (GD) method has been calling the attention of the community due to its simplicity and effectiveness \cite{Ovalle:2017fgl,Ovalle:2019qyi,Ovalle:2020kpd}. It allows one to generate new exact analytical solutions of the Einstein's equations by considering additional sources to the stress-energy tensor. Including the description of anisotropic stellar distributions \cite{daRocha:2020rda,Tello-Ortiz:2020svg}, whose predictions might be tested in astrophysical observations  \cite{daRocha:2020jdj,Fernandes-Silva:2019fez,daRocha:2017cxu,daRocha:2019pla}.  Particularly interesting for us is the recent discovery of hairy black hole solutions by gravitational decoupling. Such solution describes a black hole with hair represented by generic fields surrounding the central source of the vacuum Schwarzschild metric, requiring the existence of a well defined event horizon and that hair obeys the strong energy condition outside the horizon \cite{Ovalle:2020kpd}. Some interesting consequences of those hairy black holes has been investigated \cite{Ovalle:2021jzf,Meert:2021khi, Cavalcanti:2022cga} and much has to be done in the future. 

On the other hand, it is usually believed that a minimal length in the spacetime is related to a generalization of the uncertainty principle in the quantum realm in a plethora of theories and models \cite{Gross:1987ar,Amati:1988tn,Rovelli:1994ge,Scardigli:1999jh,HoffDaSilva:2020uov,Hossenfelder:2012jw,Tawfik:2015rva}. The link between them may be heuristically described \cite{Bishop:2022ab} by noticing that in natural units the Schwarzschild radius, $r_{\rm s}$, scales as $r_{\rm {s}}\sim M$. In higher energies, where small length scales are scrutinized, the previous relation would be transposed to $\Delta x\sim \Delta p$, so that the typical product $\Delta x \Delta p$ would have a correction proportional to $\Delta p^2$. Since {black} holes are {\it per si} physical systems under extreme conditions, their neighborhood is indeed the natural place to investigate quantum effects in the scope of the generalized uncertainty principle.
   
This paper is organized as follows: Sec. \ref{sHorizons} is dedicated to introduce basic facts about the horizon structure of the hairy black hole obtained by the gravitational decoupling procedure, obtaining three different metrics for gravitational decoupled hairy black holes. It includes an analysis of the role of the $\ell$ and $\alpha$ parameters in the resulting horizon structure. In Sec. \ref{quantueff} we investigate quantum near horizon effects of the hairy black hole, namely the correction of the Hawking radiation coming from the Dirac equation considering the generalized uncertainty principle.  Sec. \ref{4} is dedicated to conclusions.

\section{Hairy horizons and gravitational decoupling}{\label{sHorizons}}

\label{Sgd}

The extended gravitational decoupling method (EGD), introduced in \cite{Ovalle:2017fgl}, is a powerful technique for simplifying the Einstein's field equations when additional fonts are considered in a previously known seed spacetime. Departing from the Einstein's field equations,
\begin{equation}
\label{corr2}
G_{\mu\nu}
=
8\pi\,\check{T}_{\mu\nu},
\end{equation}
where $G_{\mu\nu}=
R_{\mu\nu}-\frac{1}{2}R g_{\mu\nu}$ denotes the Einstein tensor, the method assumes that the energy-momentum tensor can be split as 
\begin{equation}
\label{emt}
\check{T}_{\mu\nu}
=
T_{\mu\nu} + \Theta_{\mu\nu}.
\end{equation}
The $T_{\mu\nu}$ regards a perfect fluid that is source of a known solution of general relativity, whereas $\Theta_{\mu\nu}$ may contain new fields or an extension of the gravitational sector. The conservation equation $
\nabla_\mu\,\check{T}^{\mu\nu}=0$ must also hold. By inspecting the field equations it is possible to identify the effective density, tangential and radial pressures
\bes
\beq
\check{\rho}
&=&
\rho+
\Theta_0^{\ 0},\label{efecden}\\
\check{p}_{t}
&=&
p
-\Theta_2^{\ 2}, 
\label{efecpretan}\\\check{p}_{r}
&=&
p
-\Theta_1^{\ 1}.
\label{efecprera}
\eeq\ees

\begin{figure}[htp]
 \centering
 \begin{center}
\includegraphics[width=\textwidth]{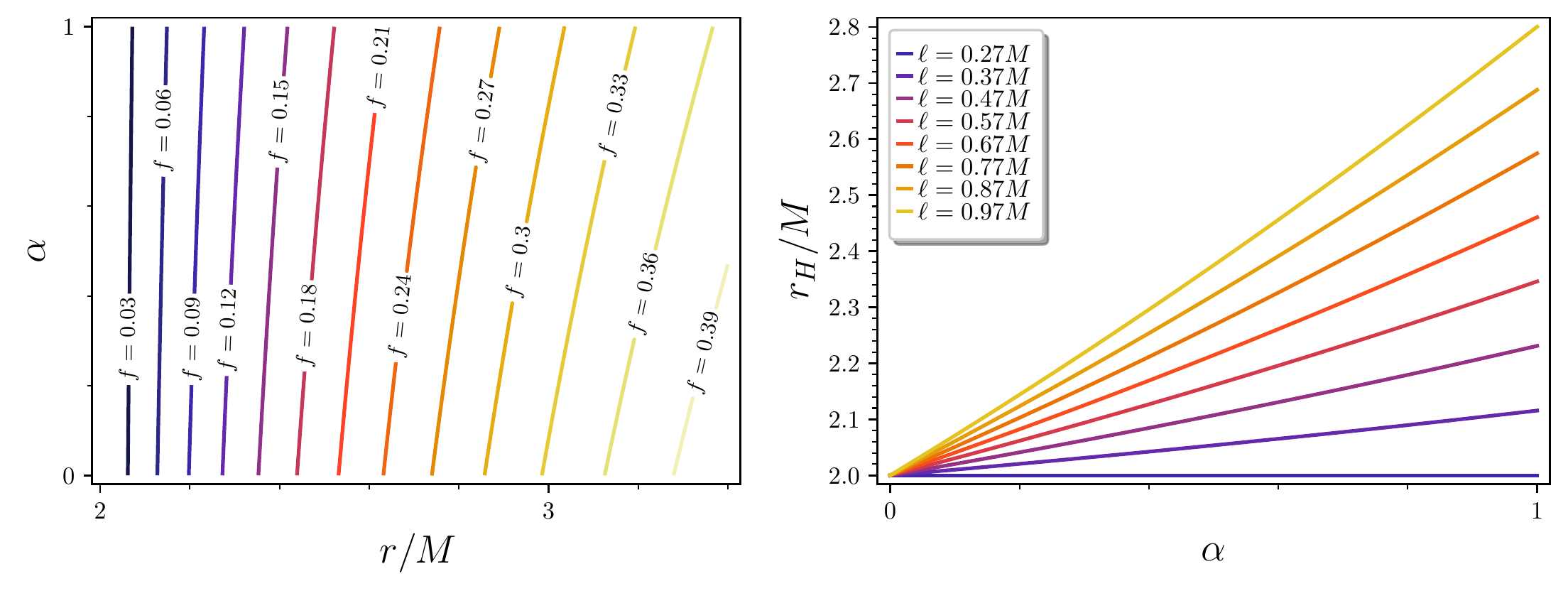}
\end{center}
 \caption{Left panel: Contour lines of $f(r)$ depending on $(r,\alpha)$ for the extreme case ($\ell=2Me^{-2}$). Right panel: Radius of the hairy black hole horizon $r_H$ as function of $\alpha$ for different values of the parameter $\ell$. The non-linearity of Eq. \eqref{eq_rh} has little influence on $r_H$.}
 \label{Fig1}
\end{figure}

The idea now is take a spherically symmetric solution to the field equations and deform it in a way such that the field equations split in a sector containing the known solution and another one for the deformation.  Indeed, assuming a known spherically symmetric metric  
\begin{equation}
ds^{2}
=
-e^{\kappa (r)}dt^{2}
+e^{\upzeta (r)}dr^{2}
+
r^{2}d\Upomega^2
,
\label{pfmetric}
\end{equation}
and deforming $\kappa(r)$ and $\upzeta(r)$ as
\bes
\begin{eqnarray}
\label{gd1}
\kappa(r)
&\mapsto &
\kappa(r)+\alpha f_2(r)
\\
\label{gd2}
e^{-\upzeta(r)} 
&\mapsto &
e^{-\upzeta(r)}+\alpha f_1(r),
\end{eqnarray}
\ees
the resulting field equations split into two distinct arrays. The first one for the source $T_{\mu\nu}$, whose solution is given by the metric~(\ref{pfmetric}). And the second encompass $\Theta_{\mu\nu}$ as well as the deformation functions $f_1(r)$ and $f_2(r)$, to be determined by the field equations. It reads
\bes
\begin{eqnarray}
\label{ec1d}
\!\!\!\!\!8\pi\,\Theta_0^{\ 0}
&=&
\alpha\left(\frac{f_1}{r^2}+\frac{f_1^\prime}{r}\right),
\\
\label{ec2d}
\!\!\!\!\!8\pi\,\Theta_1^{\ 1}
-\alpha\,\frac{e^{-\upzeta}\,f_2^\prime}{r}
&\!=\!&
\alpha\,f_1\left(\frac{1}{r^2}+\frac{\kappa'(r)+\alpha f_2'(r)}{r}\right)
\\
\label{ec3d}
\!\!\!\!\!\!\!\!\!\!\!\!\!8\pi\Theta_2^{\ 2}\!-\!\alpha{f_1}Z_1(r)\!&\!\!=\!\!&\!\!\!\alpha\frac{f_1^\prime}{4}\!\left(\kappa'(r)+\alpha f_2'(r)\!+\!\frac{2}{r}\right)\!+\!\alpha Z_2(r)
\end{eqnarray}
\ees
where \cite{Ovalle:2017fgl}
\bes
\begin{eqnarray}
Z_1(r) &=& {\alpha}^{2} f'_2\left(r\right)^{2} + 2 \,{\alpha} {\left(f'_2\left(r\right) \kappa'\left(r\right) + \frac{f'_2\left(r\right)}{r} + f''_2\left(r\right)\right)}  + \kappa'\left(r\right)^{2} + \frac{2 \, \kappa'\left(r\right)}{r} + 2 \, \kappa''\left(r\right)
\\
Z_2(r) &=& \alpha e^{-\upzeta}\left(2f_2^{\prime\prime}+f_2^{2\prime}+\frac{2f_2^\prime}{r}+2\kappa' f_2^\prime-\upzeta' f_2^\prime\right).
\end{eqnarray}
\ees
The above equations show that $\Theta_{\mu\nu}$ must vanish when the deformations parameter $\alpha$ vanish. It finishes the main setup of the gravitational decoupling procedure. In order to find a black hole solutions with a well-defined horizon structure, in Ref. \cite{Ovalle:2020kpd} the Schwarzschild solution was assumed in place of Eq. \eqref{pfmetric}, demanding that $g_{rr} = -\frac1{g_{tt}}$ for the deformed metric, namely

\beq
\left(1-\frac{2M}r\right)\left(e^{\alpha f_2(r)}-1\right) = \alpha f_1(r).
\eeq
So that 
\begin{eqnarray}
\label{hairyBH}
ds^{2}
&\!=\!&
-\left(1-\frac{2M}{r}\right)
e^{{\scalebox{.65}{${\boldsymbol\alpha}$}} f_2(r)}
dt^{2}
\!+\!\left(1-\frac{2M}{r}\right)^{-1}
e^{-{\scalebox{.65}{${\boldsymbol\alpha}$}}\,f_2(r)}
dr^2+r^{2}\,d\Upomega^2.
\end{eqnarray}
Further, assuming the strong energy conditions, 
 \bes
\begin{eqnarray}
\check{\rho}+\check{p}_r+2\,\check{p}_t
\geq
0, 
\label{strong01}\\
\check{\rho}+\check{p}_r
\geq
0,\\
\check{\rho}+\check{p}_t
\geq
0,
\end{eqnarray}
\ees
and managing the field equations, a new hairy black hole solution was found \cite{Ovalle:2020kpd}
\begin{equation}\label{metric_hairy}
	ds^2 = - f(r) dt^2 + \frac1{f(r)} dr^2 + r^2 d\Omega^2,
\end{equation}
where
\begin{equation}
f(r) = 1 - \frac{2M + \alpha\ell}{r} + \alpha e^{-\frac{r}{M}}.
\end{equation}
The dimensionless parameter ${\alpha}$ keeps track of the deformation of the Schwarzschild black hole and $\ell$, which has dimension of length, is a constant appearing as a result of a non vanishing additional font $\Theta_{\mu\nu}$. Notice that by taking $\alpha=0$ we recover the Schwarzschild solution. Also, as a result of the strong energy condition, the $\ell$ parameter is restricted to ${ \ell}
\geq
2M/e^{2}$, whose extremal case ${ \ell}=2M/e^{2}$ results 
\begin{eqnarray}
\label{strongh2M}
f(r)
=
1-\frac{2M}{r}+{\alpha}\left(e^{-\frac{r}{M}}-\frac{2M}{e^2\,r}\right).
\end{eqnarray}
The hairy black hole has only one horizon, located at $r=r_{H}$, such that 
\begin{equation} \label{eq_rh}
	r_{H} - 2M - \alpha\ell + r_{H}\alpha e^{-\frac{r_{H}}{M}}=0.
\end{equation}
Such equation has no analytical solution, except for specific values of the parameters $(\alpha,\ell)$. In particular, for the extreme case, the horizon is located at $r_{H} = 2M$. The equation \eqref{eq_rh}, however, can be analyzed numerically.  Figure \ref{Fig1} shows the horizon radius for different values of the parameters $(\alpha,\ell)$ in the range allowed by the strong energy condition.

\begin{figure}[htp]
 \centering
 \begin{center}
\includegraphics[width=.47\textwidth]{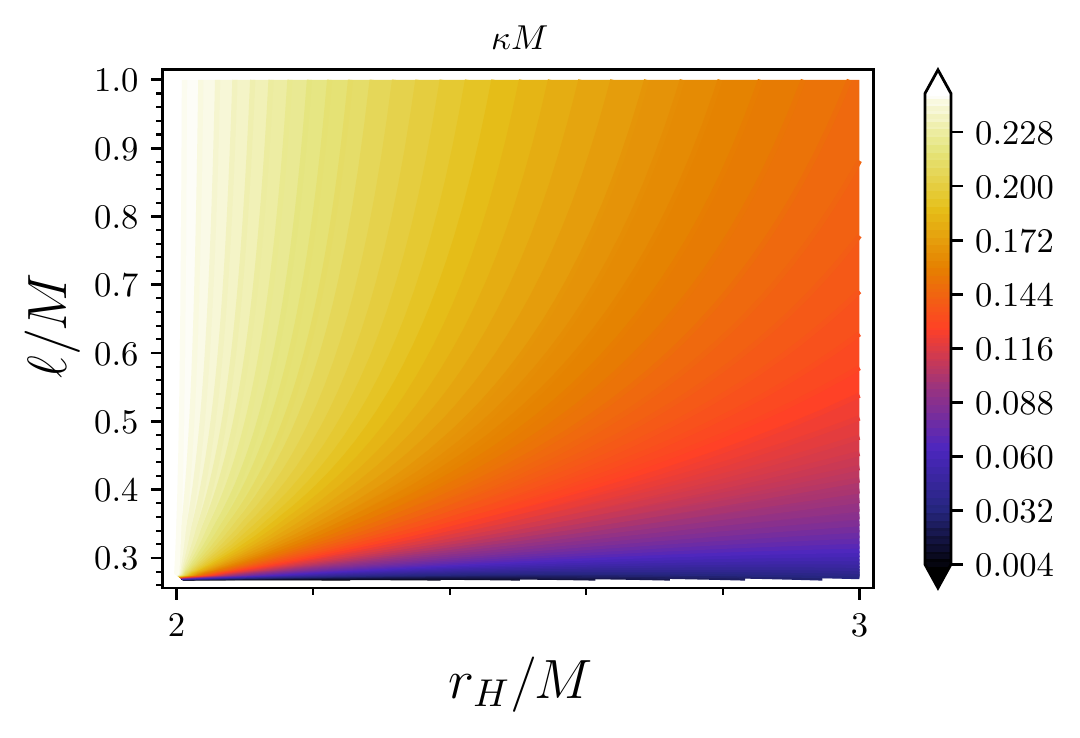}
\includegraphics[width=.47\textwidth]{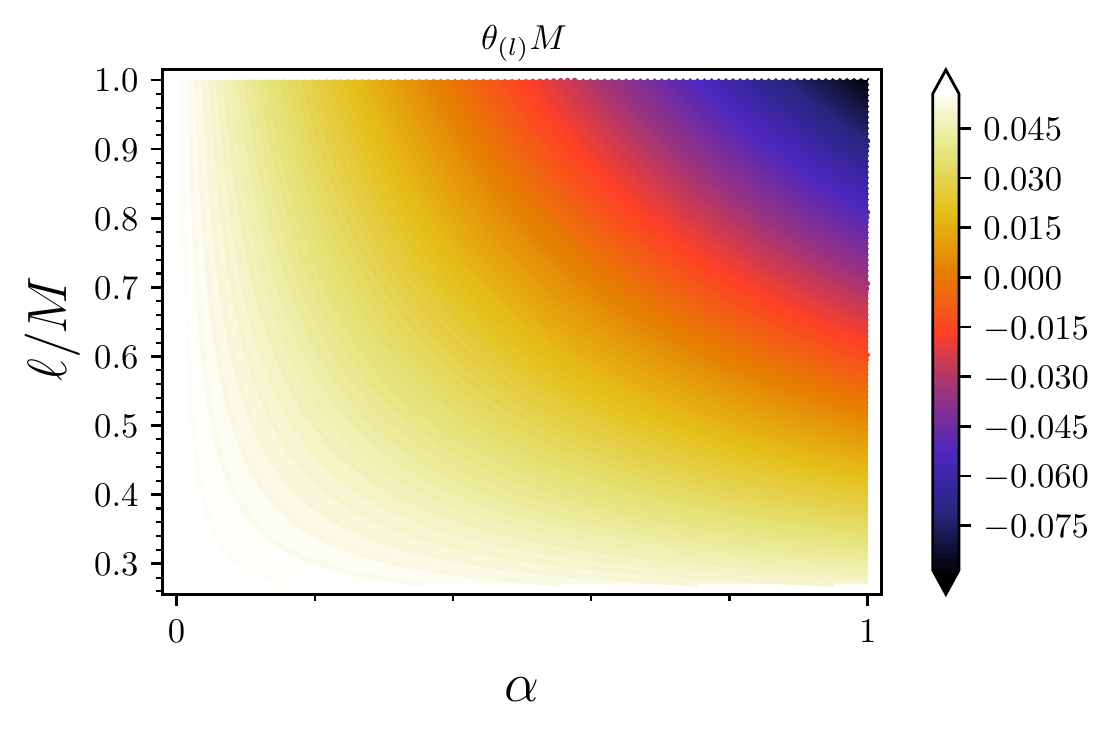}
\end{center}
 \caption{Left panel: gravitational constant $\kappa$ (color scale) as function of the horizon radius $r_H$ and hairy parameter $\ell$. The  decoupling parameters $\alpha$ was eliminated using Eq. \eqref{eq_rh}. The Schwarzschild case corresponds to the extreme left vertical line. Right panel: expansion of the cross-section (color scale) along the null vector $l$ for $r=2.3$. Notice that close to the upper right corner $r_H>2.3$ (see Figure \ref{Fig1}), resulting in a negative expansion.}
 \label{Fig2}
\end{figure}

\begin{figure}[htp]
 \centering
 \begin{center}
\includegraphics[width=.47\textwidth]{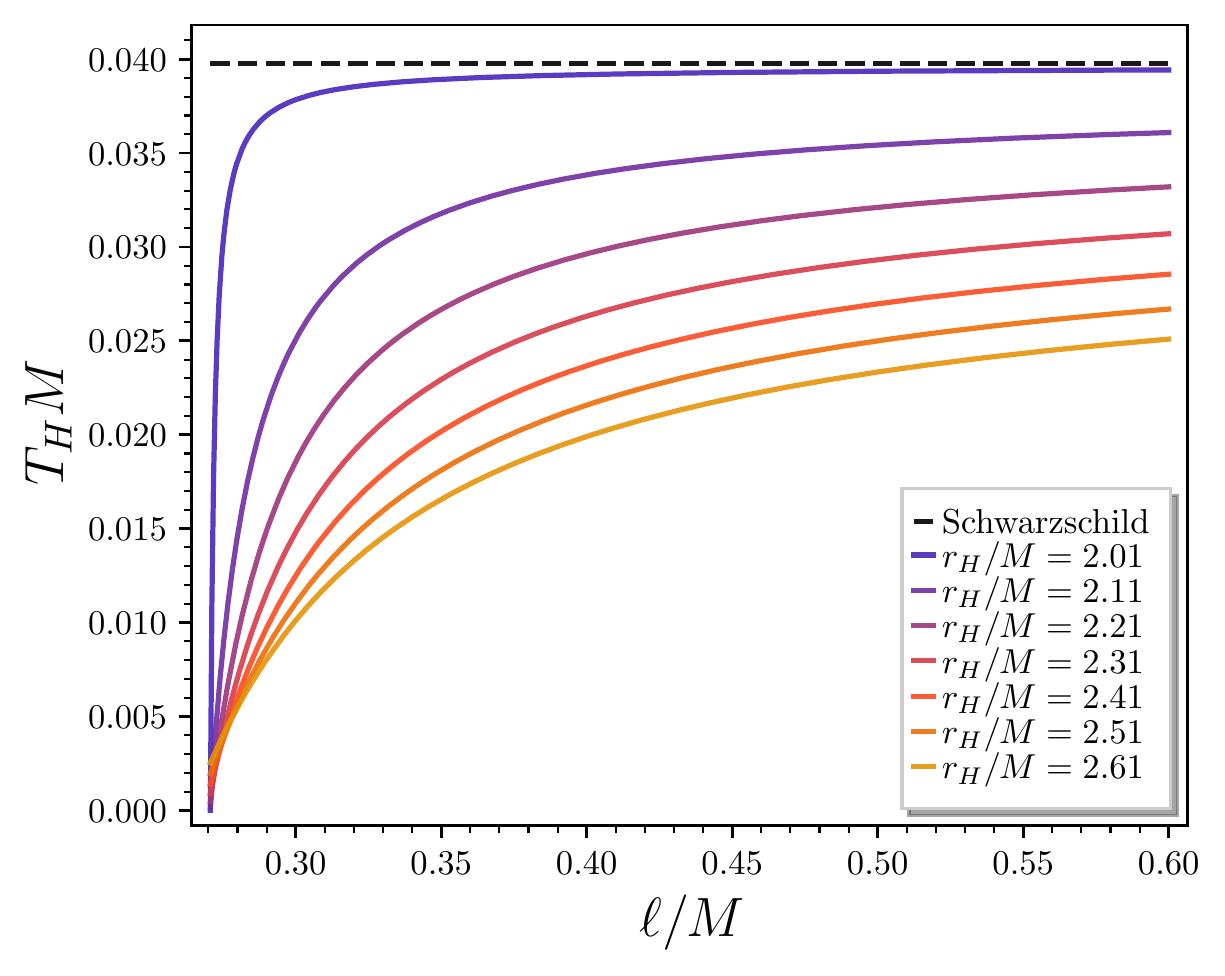}
\includegraphics[width=.47\textwidth]{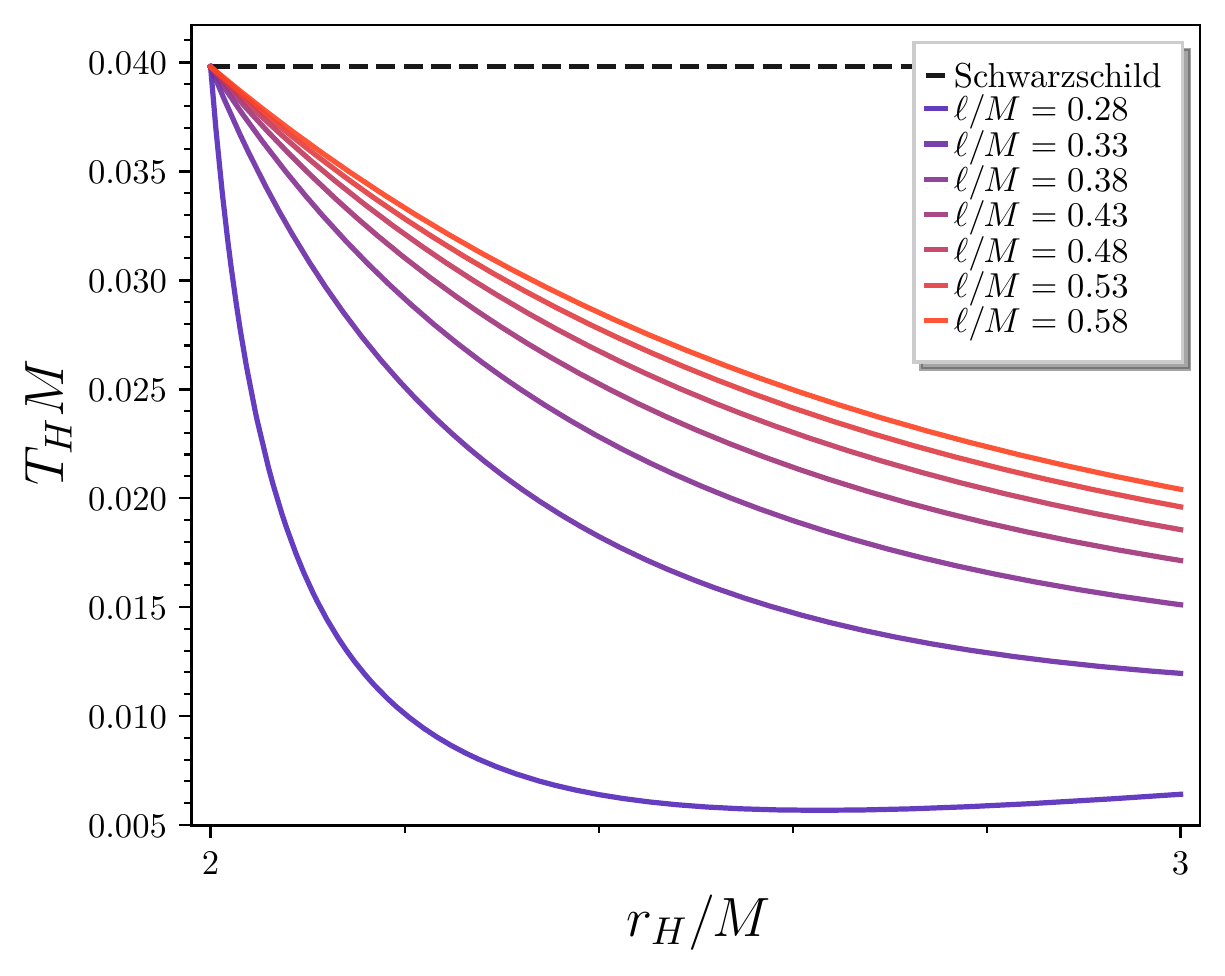}
\end{center}
 \caption{Left panel: Hawking temperature of hairy black holes depending on hairy parameter $\ell$ for different values of the horizon radius $r_H$. Notice the low temperature for $\ell \to 2Me^{-2}$. Right panel: Hawking temperature of hairy black holes depending on horizon radius $r_H$ for different values of the hairy parameter $\ell$. When $r_H \to 2M$ the temperature $T_H \to T_{\rm {Shw}}$ for any $\ell$.  In both cases the decoupling parameter $\alpha$ was eliminated using Eq. \eqref{eq_rh} and the radius $r_H = 2M$ corresponds to the extreme case. }
 \label{Fig3}
\end{figure}

In order to investigate the horizon structure of the above hairy black hole, we are going to perform a coordinate transformation to an analogous to the advanced Eddington-Finkelstein coordinates. So that $(t,r,\theta,\phi)\mapsto(v,r,\theta,\phi)$ and $v$ is given by
\begin{equation}
	 v = t - \int \left(1 - \frac{2M + \alpha \ell}{ r} + \alpha  e^{- \frac{r}{M}}\right)^{-1} dr.
\end{equation}
In those new coordinates, the metric \eqref{metric_hairy} takes the form
\begin{equation}
	ds^{2} = -\left(1 - \frac{\alpha \ell + 2 \, M}{r} +  \alpha e^{-\frac{r}{M}} \right) \mathrm{d} v^{2} +2\mathrm{d} v \mathrm{d} r + r^{2} \mathrm{d} {\Omega}^2.
\end{equation}
With the advantage of being regular on the horizon. In fact, the only physical singularity is at $ r = 0 $ \cite{Ovalle:2020kpd}, as can be seen from the Kretchmann scalar,

\begin{align}\nonumber
 K &= \frac{48 \, M^{2}}{r^{6}} - 8 \, \alpha {\left(\frac{2 \, M e^{-\frac{r}{M}}}{r^{5}} + \frac{2 \, e^{-\frac{r}{M}}}{r^{4}} + \frac{e^{-\frac{r}{M}}}{M r^{3}} - \frac{6 \, M \ell}{r^{6}}\right)} \\
 &- \alpha^{2} {\left(\frac{8 \, \ell e^{-\frac{r}{M}}}{r^{5}} + \frac{8 \, \ell e^{-\frac{r}{M}}}{M r^{4}} + \frac{4 \, \ell e^{-\frac{r}{M}}}{M^{2} r^{3}} - \frac{e^{-\frac{2r}{M}}}{M^{4}} - \frac{4 \, e^{-\frac{2r}{M}}}{r^{4}} - \frac{4 \, e^{-\frac{2r}{M}}}{M^{2} r^{2}} - \frac{12 \, \ell^{2}}{r^{6}}\right)}. 
\end{align}
We also need the normal null vectors $l$ and $k$, so that $l_{\mu}l^{\mu}=k_{\mu}k^{\mu}=0, l_{\mu}k^{\mu} = -1$ and explicitly given by
\begin{align}
	l &= \partial_{v} +\frac{1}{2} \left(1 - \frac{2M + \alpha\ell}{r} + \alpha e^{-\frac{r}{M}}\right){\partial}_{r },\\
	k &= -{\partial}_{r}.
\end{align}
They are linearly independent future-pointing and associated with the null geodesics outgoing and ingoing the horizon, respectively. It allow us to introduce metric on the horizon cross-section \cite{Gourgoulhon}
\begin{equation}
	q_{\mu\nu} = g_{\mu\nu} + l_{\mu}k_{\nu} + k_{\mu}l_{\nu} \xrightarrow{\hspace*{.7cm}} q =  r^{2} \mathrm{d}{\theta}^2 + r^{2}\sin^{2}\theta \mathrm{d}\phi^2,
\end{equation}
as well as the expansion along the null normals
\begin{equation}
	\theta_{(l)}= \frac{1}{2}\mathcal{L}_{l}\det(q) = \frac{1}{r} \left(1 - \frac{2M + \alpha\ell}{r} + \alpha e^{-\frac{r}{M}}\right),
\end{equation}

\begin{equation}
\theta_{(k)}=  \frac{1}{2}\mathcal{L}_{k}\det(q)=-\frac{2}{r},
\end{equation}
where $\mathcal{L}_{l}$ denotes the Lie derivative along $l$. As expected, on the horizon
\begin{equation}\label{expan_l_horizon}
\left.\theta_{(l)}\right|_{r=r_H}= 0,
\end{equation}
 \begin{equation}\label{expan_k_horizon}
 \left.\theta_{(k)} \right|_{r=r_H}= -\frac{2}{r_{H}}<0.
 \end{equation}
These results show that the cross section is a marginally trapped surface  \cite{Gourgoulhon:2008pu} and that the hairy horizon is a non-expanding horizon \cite{Ashtekar:2004cn} (see Figure \ref{Fig2}).
Furthermore, since on the horizon the null normal $l$ coincides with the killing vector $\xi = \partial_{v}$, it is also a Killing horizon, whose associated gravitational surface is given by
\begin{align}
\kappa = \left[\nabla_{\mu}l^{\mu} - \theta_{(l)}\right]_{r=r_H} = \frac{M}{r_H^{2}} + \frac{\alpha}{2}  {\left(\frac{\ell}{r_H^{2}} - \frac{e^{-\frac{r_H}{M}}}{M} \right)}. 
\end{align}
It straightforwardly gives the Hawking temperature of the hairy black hole,

\begin{align}
 T_H = \frac{\kappa}{2\pi} = \frac{1}{2\pi}\left[\frac{M}{r_H^{2}} + \frac{\alpha}{2}  {\left(\frac{\ell}{r_H^{2}} - \frac{e^{-\frac{r_H}{M}}}{M} \right)}\right].
\end{align}
Eliminating the decoupling parameters $\alpha$ by means of Eq. \eqref{eq_rh} we find
\begin{align}
 T_H = \frac{\left( {\ell} e^{\frac{{r_H}}{M}} - 2  M + 2  {r_H} \right)M- {r_H}^{2}}{2 \, {\left({\ell} e^{\frac{{r_H}}{M}} - {r_H}\right)} M {r_H}}.
\end{align}
For the extreme case it simplifies to
\begin{align}
 T_H =  \frac{1}{8\pi M}\left(1 - \frac{\alpha}{e^2} \right) = T_{\rm {Shw}}\left(1 - \frac{\alpha}{e^2} \right),
\end{align}
where $T_{\rm {Shw}} =  \frac{1}{8\pi M}$ is the Hawking temperature of the Schwarzschild black hole. Notice that, when compared to the Schwarzschild case, the extreme hairy black hole is slightly colder. Furthermore, for $\ell$ close to the extreme value $2Me^{-2}$, the Hawking temperature $T_H \to 0$ (see Fig. \ref{Fig3}).  In next section we extend such result to a scenario of generalized uncertainty principle (GUP) and fermionic emission as Hawking radiation.

Since the field equations comes from the standard Einstein-Hilbert action, the entropy of the hairy black hole shall be the Bekenstein-Hawking entropy,
\begin{align}
 S = \pi r_H^2.
\end{align}
In order to examine the thermal stability of the hairy black hole we should find the heat capacity at constant $\ell$, given by
\begin{align}
 C_\ell &= T\left(\dfrac{\partial S}{\partial T}\right)_\ell=T\left(\dfrac{\partial S}{\partial r_H}\right)_\ell\left(\dfrac{\partial T}{\partial r_H}\right)_\ell^{-1}\\
 &= -\frac{2 \, \pi {\left({r_H}^{2} - {\ell} e^{{r_H}} - 2 \, {r_H} + 2\right)} {r_H}^{2}}{{r_H}^{3} - 2 \, {r_H}^{2} - 2 \, {\ell} e^{{r_H}} + 4}.
\end{align}
\begin{figure}[htp]
 \centering
 \begin{center}
\includegraphics[width=.47\textwidth]{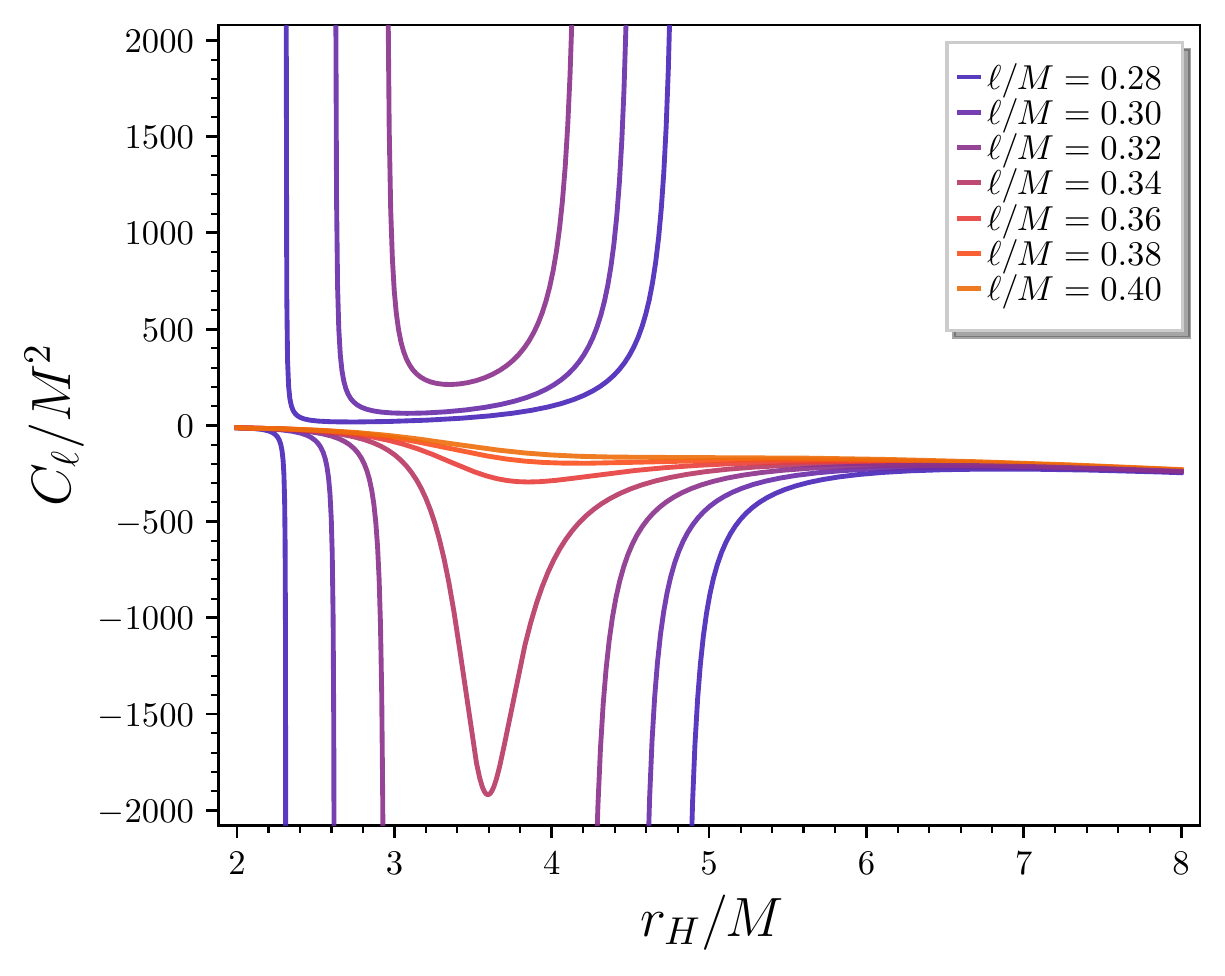}
\includegraphics[width=.47\textwidth]{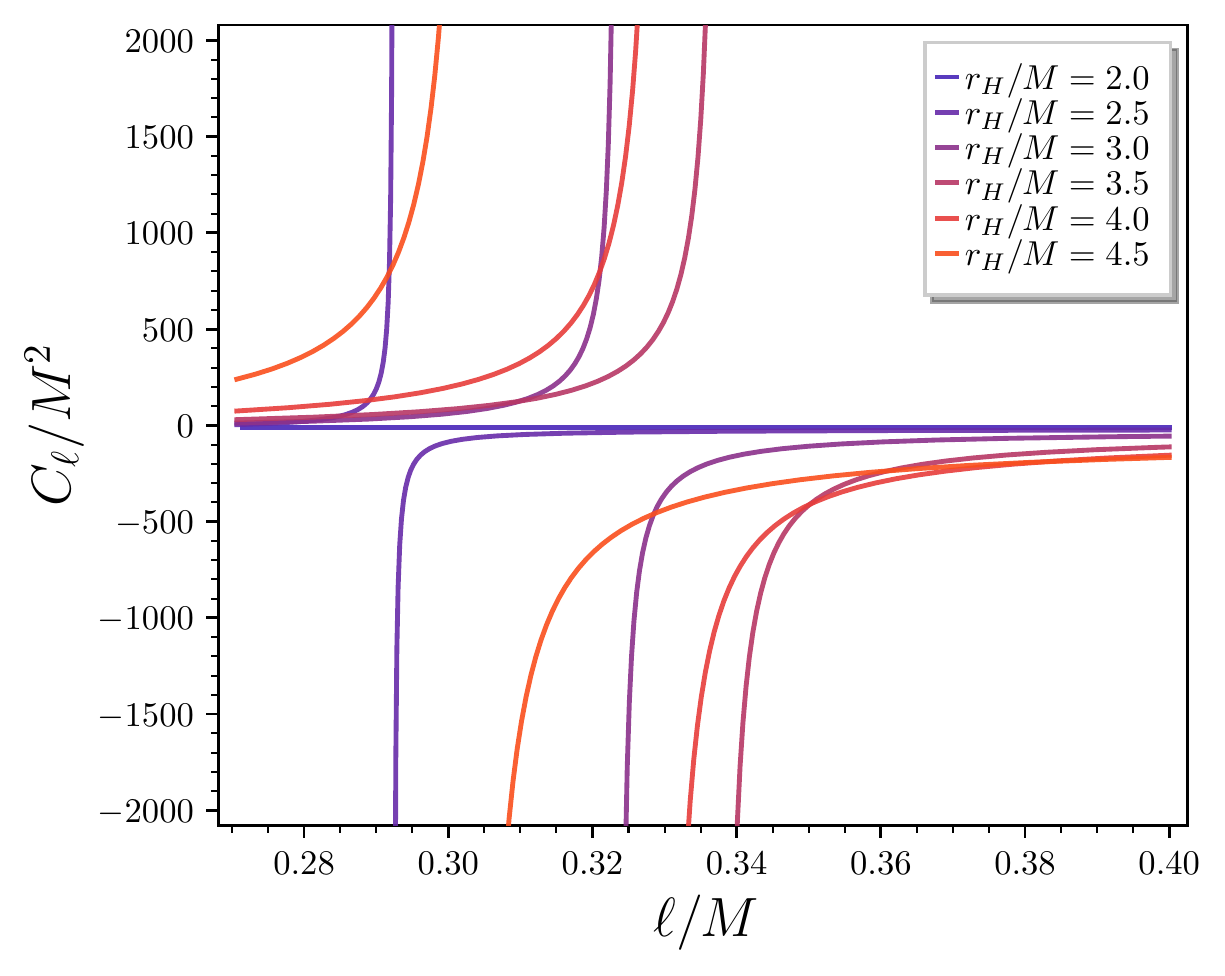}
\end{center}
 \caption{Left panel: heat capacity of hairy black holes depending on the horizon radius $r_H$ for different values of hairy parameter $\ell$. Right panel: heat capacity of hairy black holes depending on the hairy parameter $\ell$ for different values of the horizon radius $r_H$. In both cases the decoupling parameter $\alpha$ was eliminated using Eq. \eqref{eq_rh} and the radius $r_H = 2M$ corresponds to the extreme case. Notice the discontinuities for $\ell <.36$ and $r_H<5$.}
 \label{Fig4}
\end{figure}
The discontinuity of the heat capacity, as shown in Fig. \ref{Fig4}, reflects the appearance of a Hawking-Page like phase transition \cite{Hawking:1982dh}, separating regions of stable and unstable domains. Such discontinuities appear only for small $\ell$ and relatively small $r_H$, namely for $\ell \lesssim .35$ and $r_H \lesssim 5$, beyond which the heat capacity tends to saturate. The heat capacity is also well behaved when $r_H=2M$, regardless of $\ell$. Again this highlights that the exotic behaviour of the hairy black hole is manifested only for small $r_H$ and small $\ell$, but beyond $r_H=2M$ and the extreme case ($\ell = 2Me^{-2}$). Furthermore, from the heat capacity plots in Fig. \ref{Fig4}, left panel, we see that increasing $\ell$ the discontinuity points become closer and closer, eventually merging and consequently turning the discontinuity into a peak. Which in turn spreads to saturation.

\section{Quantum effects near the horizon}\label{quantueff}

\subsection{GUP and generalized Dirac equation}
Minimum lengths are predicted from different approaches to quantum gravity, as string theory \cite{Gross:1987ar,Amati:1988tn}, loop quantum gravity \cite{Rovelli:1994ge}, quantum black holes \cite{Scardigli:1999jh}, among others \cite{HoffDaSilva:2020uov,Tawfik:2015rva,Hossenfelder:2012jw}. Some of those efforts have led to the so called \textit{Generalized Uncertainty Principle} (GUP), from where a minimum length raises naturally, 

\begin{eqnarray}
\Delta x \Delta p \geq \frac{\hbar}{2}\left[1+ \beta \Delta
p^2\right], \label{eq2.1}
\end{eqnarray}
where $\beta =\beta_0/m_p^2$, $m_p$ is the Planck mass and
$\beta_0 $ is a dimensionless parameter. In order to encompass the effects coming from the GUP, in Ref. \cite{Kempf:1994su} were made modifications
on the commutation relations $\left[x_i,p_j\right]= i \hbar
\delta_{ij}\left[1+ \beta p^2\right]$, where $x_i$ and $p_i$ are
position and momentum operators defined by $x_i = x_{0i}$ and $p_i = p_{0i} (1 + \beta p^2)$ respectively, where $x_{0i}$ and $p_{0j}$ satisfy the standard commutation relations. Therefore, keeping only the first order in $\beta$ one has 
\begin{eqnarray}
p^2 &\simeq & - \hbar ^2\left[ {\partial _i \partial ^i - 2\beta \hbar
^2 \left( {\partial ^j\partial _j } \right)\left( {\partial
^i\partial _i } \right)} \right]. \label{eq2.3}
\end{eqnarray}

\noindent According to \cite{greiner2000relativistic}, quantum gravity effects engender (as a net effect) a generalized frequency, with $ E = i \hbar
\partial _0 $, given by $\tilde \omega = E( 1 - \beta E^2)$. Now, by taking into account the energy mass shell condition $ p^2 +
m^2 = E^2 $, the expression of energy in this context reads
\cite{Nozari:2012nf,greiner2000relativistic,Nozari:2005ix,Hossenfelder:2003jz}

\begin{eqnarray}
\tilde E = E[ 1 - \beta (p^2 + m^2)]. \label{eq2.5}
\end{eqnarray}

\noindent In what follows, we investigate
the radiation of spin-1/2 fermions in curved spacetime where
effects of quantum gravity are taken into account. It is carried out by means of the curved spacetime version of the generalized Dirac equation \cite{Nozari:2005ix}.  The usual one in curved spacetimes is given by
\begin{align}\label{dicurv}
\left(\;i{\hslash}\gamma^a e^{\mu}_{\;\;a} D_\mu+{m}\;\right)\Psi_{\uparrow}(t,r,\theta,\phi)=0,
\end{align}
where
\begin{align}
D_\mu=\partial_{\mu }+\frac{i}{2}{}{} \omega _{\mu }^{ab}\Sigma_{ab} \equiv \partial_{\mu } + \Omega_\mu ,
\end{align}
 with $\Sigma^{ab}=\frac{i}{4}\left[ \gamma
^{a },\gamma ^{b }\right]$; $\gamma ^{d}$ are the Clifford algebra generators for the Minkowski spacetime, $\omega _{\mu }^{ab }$ are the spin connection coefficients and $e^{\mu}_{\;\;a}$ are the  vierbein fields
\begin{align}
g^{\mu\nu}=e^{\mu}_{\;\;a}e^{\nu}_{\;\;b}\eta^{ab}.
\end{align}

\noindent We adopt the convention that lowercase latin indexes denote the vierbein flat spacetime index, whereas Greek indexes are the curved spacetime ones. To avoid confusion we are going label them as the spacetime coordinates $(t,r,\theta,\phi)$ for curved spacetime and numbers for the flat one.

Combining Eqs. \eqref{eq2.3},\eqref{eq2.5} and \eqref{dicurv}, and neglecting higher orders of $\beta$, the generalized Dirac equation in curved spacetime is found \cite{Nozari:2005ix, Chen:2013tha}

\begin{align}
-i\hslash\gamma^{0}\partial_{0}\Psi_{\uparrow}(t,r,\theta,\phi) = \left(i\hslash\gamma^{i}\partial_{i}+i\hslash\gamma^{\mu}\Omega_{\mu}+{m}\right)\left(1+\beta\hslash^{2}\partial_{j}\partial^{j}-\beta
m^{2}\right)\Psi_{\uparrow}(t,r,\theta,\phi). \label{gen_dirac}
\end{align} This equation shall be used to derive corrections to the Hawking temperature by considering the GUP.

\subsection{Corrected fermionic tunneling through hairy horizon}\label{femem}

In this section we are interested on corrected spin-$1/2$ fermions emission as Hawking radiation. Those  particles are expected to be emitted as Hawking radiation due to the fact that black holes are surrounded by a thermal bath of finite temperature, from where all sort of particles could emerge  \cite{Page:1976df,Kerner:2007rr}. 
The key point here is replacing the Dirac equation by its generalized version, as introduced in the previous section. It produces, as we are going to see, new corrections to the Hawking radiation of hairy black holes. Apart from replacing the Dirac equation, the procedure is the usual one for the tunneling method \cite{Barducci:1976qu,Vanzo:2011wq, Kerner:2007rr, Cavalcanti:2015nna}. The first point consists in choosing a spin up or spin down spinor and apply the WKB approximation. The spin up, for example, results
\begin{align}
\Psi_{\uparrow}(t,r,\theta,\phi)=\begin{pmatrix}
A(t,r,\theta,\phi)\\
0\\
B(t,r,\theta,\phi)\\
0
\end{pmatrix}\exp\left[\frac{i}{\hslash}I_{\uparrow}(t,r,\theta,\phi)\right],
\end{align}
where $A(t,r,\theta,\phi), B(t,r,\theta,\phi)$ are complex functions of the spacetime coordinates. One can thus substitute the above spinor in the Dirac equation and find the imaginary part of the action. The imaginary radial part encodes the tunnelling probability which, by setting it equals to the Boltzmann factor gives the temperature. Before proceeding, notice that by applying the operator ${\hslash}D_\mu$ to $\Psi_\uparrow$ most of the resulting terms are higher order in $\hslash$. In fact,

\begin{align}
{\hslash}D_\mu\Psi_{\uparrow}(t,r,\theta,\phi)=&\,{\hslash}\begin{pmatrix}
\partial_\mu A\\
0\\
\partial_\mu B\\
0
\end{pmatrix}
e^{\frac{i}{\hslash}I_{\uparrow}}+{i}\partial_\mu I_\uparrow \Psi_{\uparrow}-\frac{\hslash}{8}\omega^{ab}_\mu\Sigma_{ab}\Psi_{\uparrow}\\ \label{leadingdirac}
=&\,{i}\partial_\mu I_\uparrow \Psi_{\uparrow}+\mathcal{O}(\hslash).
\end{align}
Accordingly, we have to consider only the action derivative term for the usual Dirac operator. Here we are going to use the extreme case of the hairy black hole metric, given by
\begin{equation}
	ds^2 = - f(r) dt^2 + \frac1{f(r)} dr^2 + r^2 d\Omega^2
\end{equation}
with
$f(r)=1-\frac{2M}{r}+\alpha\left(e^{-\frac{r}{M}}-\frac{2M}{e^2\,r}\right)$. The vierbein fields of the hairy extreme spacetime metric, required to find $\gamma^ae^{\mu}_{\;\;a}$, are given by

\begin{align}
e^{t}_{\;\;0}=&\frac{1}{\sqrt{f(r)}}\,,\qquad e^r_{\;\;1}={\sqrt{f(r)}}\,,\\
e^{\theta}_{\;\;2}=&\frac{1}{r}\;,\qquad
e^{\phi}_{\;\;3}=\frac{1}{r\sin \theta}\,.
\end{align}
Hence, the representation of $\gamma^\sigma$ matrices are chosen accordingly
\begin{align}\label{v1}
e^{t}_{\;\;0}\gamma^0
=\frac{i}{\sqrt{f(r)}}\left(\begin{array}{cc}
\mathbb{I}_2 & 0 \\ 
0&-\mathbb{I}_2 
\end{array}\right)\,,&\quad e^r_{\;\;1}\gamma^1=
{\sqrt{f(r)}}\left(\begin{array}{cc}
0 & \sigma^3 \\ 
\sigma^3 & 0
\end{array}\right)\,,\\  \label{v2}
e^\theta_{\;\;2}\gamma^2=
\frac{1}{r}\left(\begin{array}{cc}
0 & \sigma^1 \\ 
\sigma^1 & 0
\end{array}\right)\,,&\quad
e^\phi_{\;\;3}\gamma^3=
\frac{1}{r\sin\theta}\left(\begin{array}{cc}
0 & \sigma^2 \\ 
\sigma^2 & 0
\end{array}\right)\,.
\end{align}
Substituting Eqs. \eqref{v1} and \eqref{v2} into Eq. \eqref{gen_dirac} and taking into account the leading order of $\hslash$, one finds the following system of equations

\begin{align}\label{dirac01}
-iA\frac{1}{\sqrt{f}}\partial_{t}I_\uparrow+\left(Am-B\sqrt{f}\partial_{r}I_\uparrow\right)\left(\beta\Lambda-1+\beta m^2\right)=0,\\
\label{dirac02}
iB\frac{1}{\sqrt{f}}\partial_{t}I_\uparrow+\left(Bm-A\sqrt{f}\partial_{r}I_\uparrow\right)\left(\beta\Lambda-1+\beta m^2\right)=0,\\
\label{dirac03}
A\left[ \left(\frac{\partial_{\theta}I_\uparrow}{r}+i\frac{\partial_{\phi}I_\uparrow}{r \sin\theta}\right)(\beta\Lambda -1+\beta m^2)\right] = 0,\\
\label{dirac04}
B\left[ \left(\frac{\partial_{\theta}I_\uparrow}{r}+i\frac{\partial_{\phi}I_\uparrow}{r \sin\theta}\right)(\beta\Lambda -1+\beta m^2)\right] = 0,\\
{\sqrt{f}}(\partial_{r}I_\uparrow)^2+\frac{1}{r}(\partial_{\theta}I_\uparrow)^2+\frac{1}{r\sin\theta}(\partial_{\phi}I_\uparrow)^2=\Lambda.
\end{align}
Notice that  Eqs. \eqref{dirac03} and \eqref{dirac04} are the same, regardless of $A$ and $B$. It means that the inward and outward tunneling angular equations are the same. Consequently, the contribution from $J(\theta,\phi)$ cancels out upon dividing the outcoming probability by the incoming probability \cite{Kerner:2007rr}. They shall be used, however, to simplify the system.
The spacetime symmetry motivates the ansatz 
\begin{align}\label{ansatz1}
I_\uparrow=-\omega t+W(r)+J(\theta,\phi).
\end{align}
Replacing it in Eq. \eqref{dirac03} or \eqref{dirac04} gives
\begin{align}\label{angular}
\left(\partial_\theta J(\theta,\phi)+\frac{i}{\sin\theta}\partial_\varphi J(\theta,\phi)\right)(\beta\Lambda -1+\beta m^2) = 0,
\end{align}
which implies $\partial_\theta J(\theta,\phi)+\frac{i}{\sin\theta}\partial_\varphi J(\theta,\phi)=0$, as the second term of Eq. \eqref{angular} does not vanish \cite{Chen:2013tha}. Consequently 
\begin{align}\label{constr_anglar}
 \left[\frac1r \partial_\theta J(\theta,\phi)\right]^2+\left[\frac{1}{r\sin\theta}\partial_\varphi J(\theta,\phi)\right]^2=0.
\end{align}
Using Eqs. \eqref{ansatz1} and \eqref{constr_anglar} into \eqref{dirac01} and \eqref{dirac02} yield the solution of
the radial action. Neglecting higher order terms of $\beta$ and taking $f$ near the horizon we find the particle's tunneling rate as determined by the imaginary part of the action
the radial action

\begin{eqnarray}
\textrm{Im} W_\pm (r) & = &\pm \textrm{Im} \int dr\frac{1}{f}\sqrt{m^{2}f+\omega^2}
\left(1+\beta m^{2}+\beta\frac{\omega^{2}}{f}\right) \nonumber \\
& = & \pm \pi\left(\frac{3 \, M m^{2} \beta \omega e^{2}}{\sqrt{\alpha^{2} - 2 \, \alpha e^{2} + e^{4}}} + \frac{2 \, M \omega e^{2}}{\sqrt{\alpha^{2} - 2 \, \alpha e^{2} + e^{4}}}\right),\\
& = & \mp \frac{2\pi M\omega}{1-\frac{\alpha}{e^2}}\left(1+\frac32 m^2\beta\right)
 \label{eq3.16}
\end{eqnarray}
where $W_+ [W_-]$ corresponds to outward [inward] solution. Remembering that the overall tunnelling probability is 
\begin{align}
\Gamma=\frac{\Gamma_{+}}{\Gamma_{-}}=\frac{e^{-2\mathtt{Im}I_+}}{e^{-2\mathtt{Im}I_-}}=e^{-2\mathtt{Im}(I_+-I_-)},
\end{align}
in the present case the tunneling rate
of fermions at the event horizon is
\begin{align}
\Gamma=&\frac{e^{-2\mathtt{Im}W_+ -2J(\theta,\phi)}}{e^{-2\mathtt{Im}W_- - 2J(\theta,\phi)}}=e^{-2\mathtt{Im}(W_+-W_-)},\\
 = & \exp\left[ \frac{8\pi M\omega}{1-\frac{\alpha}{e^2}}\left(1+\frac32 m^2\beta\right)\right].
\end{align}
This is the Boltzmann factor for an object with the
effective temperature

\begin{eqnarray}
T_H = \frac{1}{8\pi M}\frac{1-\frac{\alpha}{e^2}}{1+\frac32 m^2\beta} \simeq T_{\rm {Shw}}\left(1-\frac{\alpha}{e^2}\right)\left(1-\frac32 m^2\beta\right)  ,
\label{eq3.19}
\end{eqnarray}

\noindent Apart from the hairy parameter $\alpha$, the quantum effects coming from GUP explicitly reduces the temperature during the evaporation process. It agrees with previous investigations of remanents of black holes \cite{Nozari:2012nf, Chen:2013tha, Casadio:2017sze}. In this picture a black holes ceases to radiate when approaching the Planck scale while its effective temperature reaches a maximum value, leaving a remanent black hole \cite{Nozari:2012nf}. The combined effects of the deformation parameter $\alpha$ and quantum parameter $\beta$ strengthen the hypothesis of a vanishing Hawking emission. {It is also curious to notice the existence of a fine tuning between both, GUP and deformation, parameters, namely $\beta=-{2  \alpha }/{3 \, m^{2}e^{2}}$, whose net effect is cancel out both contributions to the Hawking temperature, restoring the standard Schwarzschild one.}

\section{Conclusions}
\label{4}

{
 In this paper were investigated some classical and semi-classical effects happening near the horizon of a recently discovered class of hairy black holes. Such black holes were derived by applying the gravitational decoupling technique \cite{Ovalle:2020kpd}. In particular, the role of the hairy parameters was analysed on the cross-section expansion along null normals, the surface gravitational constant, Hawking radiation, thermodynamics stability as well as the generalized Hawking radiation derived from the generalized uncertainty principle. For the later we applied tunelling method to the generalized Dirac equation. It shows that, apart from the severe attenuation caused by the presence of hair, the quantum parameter $\beta$ proceeds the suppression. It strengthen the hypothesis of a remnant after a vanishing Hawking emission, as explored in \cite{Nozari:2012nf, Chen:2013tha, Casadio:2017sze}. Such effects are important theoretical and phenomenological features of black holes, but unfortunately not observationally accessible at the present date. Our results also show that the exotic behavior of hairy black holes happen for the choice of parameters close to, but not equal to, the extreme case ($\ell=2Me^{-2}$), and horizon radius close to $r_H=2M$. It could be further explorer when searching for effects with observational signature.  Another intriguing possibility appearing in our results is that the quantum $\beta$ and deformation $\alpha$ parameters combined effects allow the existence of a fine tuning, namely $\beta=-{2  \alpha }/{3 \, m^{2}e^{2}}$, which prevents the Hawking temperature to deviates from $\frac{1}{8\pi M}$, corresponding to the temperature of the Schwarzschild black hole. } 

\subsection*{Acknowledgements}

K.S.A. is greatful to the Coordena\c{c}\~ao de Aperfei\c{c}oamento de Pessoal de N\'ivel Superior - Brasil (CAPES), Finance Code 001 and grant No 2021/03625-8, S\~ao Paulo Research Foundation (FAPESP) for financial support. J.M.H.S thanks to CNPq (grant No. 303561/2018-1) for financial support.

\bibliography{bib_hairy_horizon}

\end{document}